\documentclass[twocolumn,showpacs,preprintnumbers,amsmath,amssymb]{revtex4}
\usepackage{graphicx}% Include figure files
\begin{document}

\title{Inserting two atoms into a single optical micropotential}

\author{Y.~Miroshnychenko, W.~Alt,  I.~Dotsenko, L.~F\"{o}rster,
M.~Khudaverdyan, D.~Meschede, S.~Reick,}

\author{A. Rauschenbeutel}%
\email{rauschenbeutel@iap.uni-bonn.de}
\affiliation{%
Institut f\"{u}r Angewandte Physik, Universit\"{a}t Bonn,
Wegelerstr.~8, D-53115 Bonn, Germany }

\date{\today}% It is always \today, today,

\begin{abstract}
We recently demonstrated that strings of trapped atoms inside a
standing wave optical dipole trap can be rearranged using optical
tweezers [Y.~Miroshnychenko {\it et al.}, Nature, in press
(2006)]. This technique allows us to actively set the interatomic
separations on the scale of the individual trapping potential
wells. Here, we use such a distance-control operation to insert
two atoms into the same potential well. The detected success rate
of this manipulation is $16_{-3}^{+4}~\%$, in agreement with the
predictions of a theoretical model based on our independently
determined experimental parameters.
\end{abstract}

% PACS, the Physics and Astronomy Classification Scheme.

%32.80.Lg    Mechanical effects of light on atoms, molecules, and ions
%32.80.Pj    Optical cooling of atoms; trapping
%39.25.+k    Atom manipulation (scanning probe microscopy, laser cooling, etc.)
%03.67.-a    Quantum information

\maketitle

Controlled interaction between pairs of neutral atoms in optical
micropotentials leads to a number of interesting applications
ranging from the highly efficient production of ultracold
molecules \cite{Rom04,Xu05,Ryu05,Stoeferle06,Thalhammer06,Volz06}
to the coherent interaction of atoms through controlled cold
collisions. Such collisions have been shown to yield
state-dependent collisional phase shifts \cite{Mandel03} and to
lead to coherent spin dynamics \cite{Widera05}. Both effects are
candidates for creating entanglement and for realizing coherent
conditional dynamics, of great relevance in quantum information
processing (QIP).

So far, these experiments were carried out with large samples of
ultracold or quantum degenerate atoms, transferred into the
motional ground state of optical lattices \cite{Bloch05}. In
combination with the high atomic densities, this results in
excellent starting conditions for the above schemes, albeit at the
expense of the lack of addressability at the single atom level.
While ensemble measurements still yield information about
processes like, e.g., the entangling and disentangling dynamics
\cite{Mandel03}, their use for those QIP applications requiring
the measurement of individual quantum states is impaired.

In our ``bottom--up'' approach, on the other hand, neutral atom
systems are built atom-by-atom while maintaining full control over
the degrees of freedom of each individual atom. When stored in a
standing wave dipole trap, formed by a pair of counterpropagating
laser beams, the absolute positions of individual atoms along the
beam axis can be optically measured with sub-micrometer precision
and the number of potential wells separating simultaneously
trapped atoms can be exactly determined \cite{Dotsenko05}. In
addition, we have demonstrated that the quantum state of
individual atoms in the standing wave dipole trap can be
selectively prepared and read out with a high spatial resolution
\cite{Schrader04}. Furthermore, the atoms can be positioned along
the trap axis using the dipole trap as an ``optical conveyor
belt'' \cite{Dotsenko05,Kuhr01}. Finally, using optical tweezers,
we have recently rearranged the so far irregularly spaced atoms
into regularly spaced strings \cite{Miroshnychenko06}.

Here, we present first results concerning the insertion and
controlled interaction of two individual atoms inside the same
optical micropotential. The atoms are initially stored in separate
potential wells of a standing wave dipole trap. One of the two
atoms is then extracted out of its potential well using optical
tweezers and inserted into the potential well of the second atom.

In the rearrangement experiment \cite{Miroshnychenko06}, the final
interatomic distances were verified by recording fluorescence
images. When the two atoms ideally end up confined in a volume of
the order of one cubic optical wavelength, however, they cannot be
optically resolved. The successful insertion of the two atoms into
one potential well is therefore detected by irradiating the atoms
with near resonant light, inducing inelastic collisions. These
collisions lead to a loss of the atoms and occur if and only if
the atoms occupy the same potential well.

\begin{figure}
\includegraphics[scale=0.9]{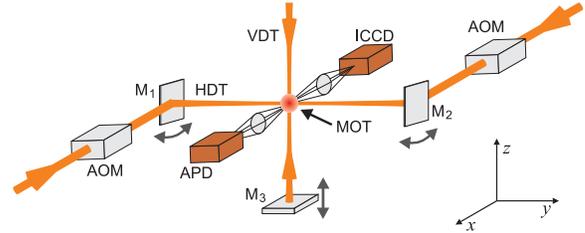}
\caption{\label{fig:setup} (Color online) Scheme of our
experimental setup. Two crossed standing wave dipole traps are
used to rearrange pairs of trapped neutral atoms. See text for
details.}
\end{figure}

The essential parts of our setup are schematically depicted in
Fig.~\ref{fig:setup}. A horizontal standing wave dipole trap (HDT)
is formed by two counterpropagating Nd:YAG laser beams with a
wavelength of $\lambda_{\mathrm{HDT}}=1064~\mathrm{nm}$ and a
power of $1~\mathrm{W}$ each. They are focused to a waist of
$w_\mathrm{HDT}=19~\mathrm{\mu m} $ ($1/e^2$-radius), generating a
chain of potential wells, separated by
$\lambda_{\mathrm{HDT}}/2=532~\mathrm{nm}$ with a measured depth
of $U_{\mathrm{HDT}} / k_{\mathrm{B}}=0.8~\mathrm{m K}$. The HDT
is loaded with an exactly known number of caesium atoms from a
high gradient magneto-optical trap (MOT), as inferred from the
discrete MOT fluorescence levels, recorded with an avalanche
photodiode (APD). The single atom transfer efficiency between the
traps is $98.7_{-1.1}^{+0.7}~\%$. Following the transfer from the
MOT, we let the atoms freely expand along the DT axis by switching
off one of the beams for 1~ms. The atoms are then randomly
distributed over an interval of about $80~\mathrm{\mu m}$ along
the axis of the trap. Subsequently, we record a fluorescence image
using an intensified CCD camera (ICCD) \cite{Miroshnychenko03}.
For this purpose, we illuminate the atoms with a near resonant
three-dimensional optical molasses, thereby also cooling the atoms
to a temperature of about $80~\mathrm{\mu K}$. From the ICCD image
with 1~s exposure time, the position of all optically resolved
atoms are determined with an uncertainty of $\Delta
y_{\mathrm{position}}=140~\mathrm{n m}$ rms along the axis of the
HDT, significantly smaller than the 532~nm separation between
adjacent potential wells of the HDT \cite{Dotsenko05}.

The atoms in the HDT can be moved along the $x$- and
$y$-directions, see Fig.~\ref{fig:setup}. Transport along the HDT
axis, i.e., the $y$-direction, is achieved by means of our
``optical conveyor belt'' method \cite{Kuhr01,Schrader01}. For
this purpose, acousto-optic modulators (AOM) mutually detune the
laser frequencies. The moving standing wave pattern thus
transports the atoms over distances of up to a few millimeters
with submicrometer precision \cite{Dotsenko05} within a few
hundred microseconds.

Displacing the HDT in the $x$-direction, i.e., perpendicular to
its axis, is realized by synchronously tilting the mirrors M$_1$
and M$_2$ in opposite directions about the $z$-axis using PZT
actuators. For small tilts of $\sim 0.1~\mathrm{mrad}$, the
modification of the interference pattern of the HDT at the
position of the atoms is negligible. With this method, trapped
atoms can be moved in the $x$-direction by up to 40~$\mu$m, i.e.,
twice the waist radius of the HDT, with a precision of a few
micrometers within 50~ms. The storage time of the atoms in the HDT
is about $8~\mathrm{s}$, limited by heating effects caused by the
phase noise of the dual-frequency synthesizer driving the AOMs.

In order to actively control the interatomic separations, we use a
second, vertical standing wave dipole trap (VDT), operated as
optical tweezers, see Fig.~\ref{fig:setup}. The VDT is generated
by an Yb:YAG laser beam ($\lambda_\mathrm{VDT}=1030$~nm), focused
to a waist of $w_{\mathrm{VDT}}=10~\mathrm{\mu m}$ at the position
of the HDT. The standing wave is produced by retro-reflecting the
beam with a spherical mirror M$_3$. A typical incident power of
%$0.33~\mathrm{W} $
$0.3~\mathrm{W}$ results in a measured trap depth of
$U_{\mathrm{VDT}} /
k_{\mathrm{B}}=1.5~\mathrm{mK}$. The retro-reflecting mirror %M$_3$
is mounted on a PZT stage, allowing us to move the standing wave
pattern along the VDT axis. We thereby transport atoms in the
$z$-direction by typically 60~$\mu$m with a precision of a few
micrometers within 30~ms. The storage time in the VDT is about
$13~\mathrm{s}$ and is limited by heating effects caused by the
laser intensity noise of the Yb:YAG-laser. When cooling the atoms
with the optical molasses, the storage time in each of the traps
can be increased to about 1~min, limited by background gas
collisions only.

\begin{figure}
\includegraphics[scale=0.28]{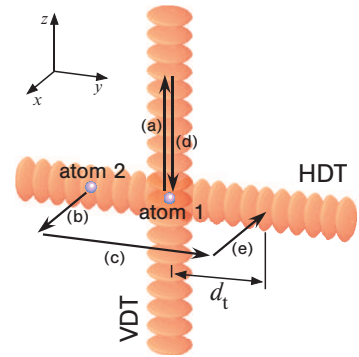}
\caption{\label{fig:radialinsertion} (Color online) The distance
between two atoms which are simultaneously trapped in the
horizontally oriented standing wave dipole trap (HDT) can be set
to a target distance $d_\mathrm{t}$ by our distance-control
operation, involving displacements of the atoms in all three
spatial dimensions. See text for details. }
\end{figure}

In both standing wave dipole traps the potential wells are almost
two orders of magnitude tighter in the axial direction than in the
radial direction. The maximum axial confining forces are thus much
larger than the maximum radial forces. As a consequence, an atom
stored in the overlap region of both traps will always follow the
axial motion of the traps. This allows us to actively set the
distance between the two atoms in the HDT: Atom~1 is first
transported along the $y$-direction into the overlap region of
both traps. Then, the standing wave pattern of the VDT is axially
shifted upwards and atom~1 moves in the positive $z$-direction by
about $3 w_\mathrm{HDT}$, see Fig.~\ref{fig:radialinsertion}(a).
At this separation, the HDT exerts negligible forces on atom~1.
Atom~2 can now be transported to any position along the HDT with
respect to the VDT even when shuttling it through the VDT. By
reinserting atom~1 into the HDT, it can hence be placed at any
target position relative to atom~2.

The reinsertion of atom~1 into the HDT is however non-trivial, if
the target distance to atom~2 is smaller than the waist of the
VDT. In this case, reinserting atom~1 by transporting it along the
VDT axis would inevitably expel atom~2 downwards out of the
overlap region. We circumvent this problem using the procedure
schematically depicted in Fig.~\ref{fig:radialinsertion}: The two
traps are first horizontally separated by displacing the axis of
the HDT in the positive $x$-direction (b). Atom~2 is then
transported to the desired $y$-position with respect to the VDT
(c), and atom~1 is transported downwards to the vertical
$z$-position of the horizontal trap (d). Next, atom~1 is inserted
at the desired position by displacing the HDT radially to the
$x$-position of the VDT (e). Finally, the VDT is adiabatically
switched off. As a result, atom~2 is not expelled out of the
overlap region, because no axial motion of the traps is involved
when merging them. Therefore, the ``radial reinsertion'' is
compatible with the insertion of atom~1 into the potential well of
the HDT already occupied by atom~2.

We first characterize the performance of our distance-control
operation for a non-zero target distance of $d_\mathrm{t}=15.00\
\mu$m between the atoms \cite{Miroshnychenko06}. For this purpose,
we load two atoms on average into the HDT and post-select the
events with initially two atoms. We use the radial insertion
scheme, as depicted in Fig.~\ref{fig:radialinsertion}, with the
order of steps (b) and (c) interchanged. The whole procedure,
including the initial ICCD image, takes about 2~s, short compared
to the trap storage times. The final distance is then checked by
recording another ICCD image.
\begin{figure}
\includegraphics[scale=0.9]{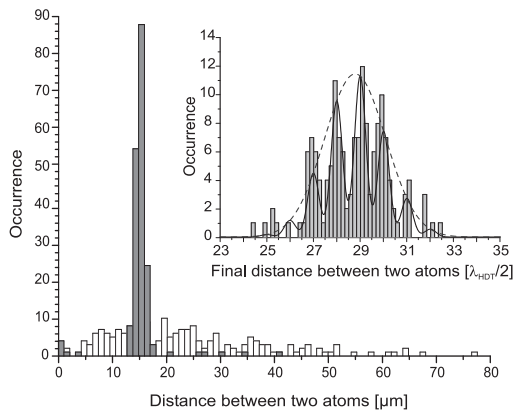}
\caption{\label{fig:dchistogram} %(Color online)
Performance of our distance-control operation. The white histogram
shows the broad distribution of the initial atomic separations for
about 190 atom pairs. The grey histogram shows the distribution of
final distances for the same pairs after the distance-control
operation (target distance $d_\mathrm{t}=15~\mathrm{\mu m}$).
Inset: Zoom of the distribution of final distances. The histogram
clearly shows that the final distances are integer multiples of
the standing wave period of $\lambda_\mathrm{HDT}/2$. The solid
line is a theoretical fit with a Gaussian envelope (dashed line)
centred at $d_\mathrm{Gauss}=15.31(\pm 0.07)~\mu$m and having a
$1/\sqrt e$-halfwidth of $\Delta d_\mathrm{Gauss}=0.71(\pm
0.05)~\mu$m. The narrow peaks under this envelope have a $1/\sqrt
e$-halfwidth of $\Delta d_\mathrm{ICCD}=0.130(\pm 0.010)~\mu$m,
corresponding to the precision of our distance measurement.}
\end{figure}
Figure~\ref{fig:dchistogram} shows a histogram (bin size of
$\lambda_\mathrm{HDT}=1064$~nm) of the initial distances (white)
and final distances (grey) between the atoms for about 190 pairs.
Initially, the atoms have random separations of up to $\sim
80~\mu$m whereas the final distribution is strongly peaked around
$d_\mathrm{m}=15.27(\pm 0.05)~\mathrm{\mu m}$. Those atom pairs
having an initial separation exceeding $10~\mu$m, i.e., the
``size'' of the optical tweezers, are rearranged by our
distance-control operation with a success rate of up to
$98_{-5}^{+2}~\%$ and a standard deviation of $\Delta
d_\mathrm{m}=0.78(\pm0.05)~\mathrm{\mu m}$
\cite{Miroshnychenko06}. This spread is mainly due to the
precision of the transport of the atoms along the HDT and the
accuracy of reinserting atom~1 into the HDT. In previous work we
have shown that our transport along the HDT is subject to a
statistical error $\Delta y_\mathrm{transp} =
0.190(\pm0.025)~\mathrm{\mu m}$~rms \cite{Dotsenko05}. Since the
experimental sequence used here involves two transports along the
HDT, i.e., moving the atom to be extracted to the $y$-position of
the VDT and then placing the remaining atoms at the target
distance $d_\mathrm{t}$, this effect contributes an uncertainty of
$\sqrt 2 \Delta y_\mathrm{transp} = 0.270(\pm0.035)~\mathrm{\mu
m}$~rms to the final distance between the atoms. Furthermore,
immediately after reinserting the extracted atom into the HDT, its
measured position has a spread of $\Delta
y_\mathrm{insert}=0.65(\pm0.05)~\mathrm{\mu m}$~rms \cite{Budget}.
Finally, the distance measurement contributes an uncertainty of
$\Delta d_\mathrm{ICCD}=0.130(\pm0.010)~\mathrm{\mu m}$. The total
expected uncertainty of the measured final distance thus amounts
to $(2\Delta y_{\mathrm{transp}}^2+\Delta
y_{\mathrm{insert}}^2+\Delta d_\mathrm{ICCD}^2)^{1/2}=0.72(\pm
0.05)~\mathrm{\mu m}$.

The fact that the distribution of final distances extends over
only a few potential wells of the HDT is strikingly apparent in
the inset of Fig.~\ref{fig:dchistogram}, where the histogram of
the distribution of final distances is displayed for a smaller bin
size of $\lambda_\mathrm{HDT}/12=89$~nm. The distribution is
clearly peaked with a periodicity of 532~nm, showing that the
final distances are integer multiples of the standing wave period
$\lambda_\mathrm{HDT}/2$. Given the width of the distribution of
true final distances $\Delta d_\mathrm{true}=(\Delta
d_\mathrm{Gauss}^2-\Delta
d_\mathrm{ICCD}^2)^{1/2}=0.70(\pm0.05)~\mathrm{\mu m}$, we can
estimate the success rate of preparing pairs of atoms separated by
a predefined number of potential wells to equal
\begin{equation}\label{eq:p_theor}
    p_{\mathrm{theor}}= \frac{p_\mathrm{noloss}}{\sqrt{2 \pi}\Delta
    d_\mathrm{true}} \int_{-\frac{\lambda_{\mathrm{HDT}}}{4}}
    ^{\frac{\lambda_{\mathrm{HDT}}}{4}}
    \!\!\!\exp\left(-\frac{y^2}{2\,\Delta
    d_\mathrm{true}^2}\right)dy\, ,
\end{equation}
where $p_\mathrm{noloss}$ is the probability for not losing an
atom during the manipulation. Assuming $p_\mathrm{noloss}=1$ for
the moment, we obtain $p_\mathrm{theor}= 30\pm 2\ \%$. In
particular, it should be possible to join the two atoms in one and
the same potential with a comparable success rate.

The experimental sequence realizing this situation corresponds to
the one depicted in Fig.~\ref{fig:radialinsertion} with
$d_\mathrm{t}$ set to zero. Again, the necessary condition for
selectively extracting atom~1 from the HDT is that both atoms be
initially separated by more than the 10-$\mu$m resolution of the
optical tweezers. We post-select these events by analysing the
initial ICCD fluorescence images. Finally, we discriminate events
where atom~1 has successfully been transferred into the potential
well containing atom~2 from events where the two atoms occupy
different potential wells by inducing two-atom losses. This is
achieved by illuminating the atoms with the optical molasses for
$1~\mathrm{s}$. It has been shown that radiative escape is the
leading physical mechanism for light induced collisions under
these conditions \cite{Kuppens00}. The resulting energy release
causes both atoms to leave the trap in most cases. If, on the
other hand, the atoms reside in different potential wells,
radiative escape is not possible and the atoms remain trapped.
Detecting the absence of the pair of atoms after the optical
molasses stage therefore confirms the successful joining of the
two atoms in one potential well of the HDT.

\begin{figure}
\includegraphics[scale=0.9]{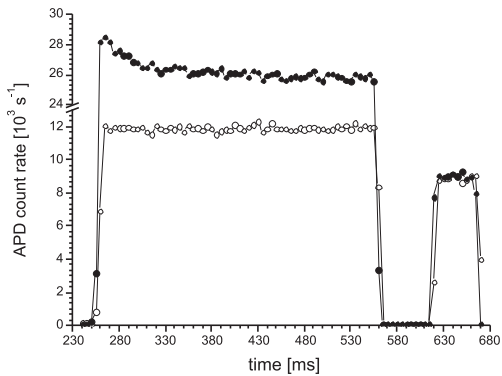}
\caption{\label{fig:fluorescence} %(Color online)
Fluorescence signal of on average 3 atoms (open circles) and on
average 19 atoms (full circles) trapped in about 25 potential
wells of the HDT. At $t=260$~ms the optical molasses illuminating
the atoms is switched on. At $t=560$~ms, the remaining atoms are
ejected from the HDT by switching off all lasers for 50~ms in
order to measure the background signal due to stray light. Each of
the two traces is averaged over 100 shots. See text for details.}
\end{figure}

In order to independently examine the dynamics of this collisional
process, we load a variable number of atoms from the MOT into the
HDT, illuminate them with the optical molasses, and detect the
atomic fluorescence with the APD. The level of this fluorescence
signal is a measure of the number of trapped atoms. If we load on
average $3$ atoms per shot distributed over about 25~potential
wells into the HDT, the probability for having two atoms in one
potential well is negligibly small. In this case, their
fluorescence level remains constant, i.e.,  no atom losses are
detected, see open circles in Fig.~\ref{fig:fluorescence}. For on
average $19$ atoms per shot distributed over about 25~potential
wells, however, the probability for at least two atoms occupying a
common well is significant. In this case, we observe an
exponential decay of the average fluorescence level to a steady
state value within about $150~\mathrm{ms}$, i.e., the optical
molasses results in radiative escape of atoms within a few hundred
milliseconds, see filled circles in Fig.~\ref{fig:fluorescence}.
The steady state fluorescence level then corresponds to the atoms
which are trapped in individual potential wells of the HDT. Our
choice of 1~s illumination time thus ensures that all pairs of
atoms undergo a light induced collision.

Since we cannot distinguish a two-atom loss due to radiative
escape from two uncorrelated one-atom losses during the
experimental sequence, we need to quantify the latter in an
independent measurement. For this purpose, we have carried out the
entire experimental sequence with only atom~1 present and only
atom~2 present. In both cases, we have measured the loss
probability $p_i$ of atom~$i$, yielding $p_1=6.5_{-2.4}^{+2.1}~\%$
and $p_2=0.0_{-0.0}^{+3.5}~\%$, respectively. From these
measurements, we infer the probability for two uncorrelated
one-atom losses during the experimental sequence to be
$p_{\mathrm{uncorr}}=p_1\cdot p_2=0.0_{-0.0}^{+0.2}~\%$.
Furthermore, the probability for not losing any of the two atoms
during the manipulation is
$p_\mathrm{noloss}=(1-p_{\mathrm{1}})(1-
p_{\mathrm{2}})=94_{-5}^{+6}~\%$. In the present measurement,
$\Delta y_{\mathrm{insert}}=0.82(\pm0.11)~\mathrm{\mu m} $~rms,
yielding $\Delta d_\mathrm{true} =0.86(\pm0.11)~\mathrm{\mu m}$.
According to Eq.~(\ref{eq:p_theor}), the probability for
successfully inserting atom~1 into the potential well of atom~2
should thus ideally be $p_{\mathrm{theor}}=23_{-3}^{+3} ~\%$

Carrying out the experimental sequence with both atom~1 {\em and}
atom~2, we measure a total two-atom loss probability of
$p_\mathrm{meas}=16_{-3}^{+4}~\%$. Compared with this value, the
probability of uncorrelated two-atom losses  $p_{\mathrm{uncorr}}$
is negligible, proving the successful joining of the two atoms in
one potential well. Note that, in former work, we have found
experimental evidence that light-induced collisions can also lead
to one-atom losses \cite{Ueberholz02}. Taking this effect into
consideration, the true success rate might then even be higher
than $p_\mathrm{meas}$.

Summarizing, we have inserted two atoms into a single potential
well of a standing wave optical dipole trap and we have
deterministically induced interactions between these atoms leading
to light-induced collisions. The presented results open the route
towards fascinating experiments. In particular, using
photoassociation techniques, it should become possible to build a
single ultracold diatomic molecule from its constituents and to
store and to manipulate this molecule inside our standing wave
dipole trap. Furthermore, by exploiting coherent spin-changing
collisions between two atoms trapped inside the same potential
well, one might be able to prepare an entangled Bell pair of atoms
which could then be used as a resource for quantum information
processing schemes.

We acknowledge valuable discussions with M.~Karski. This work was
supported by the DFG and the EC. I.~D.~acknowledges funding by
INTAS.


\begin{thebibliography}{99}
\bibitem{Budget}
The main contributions to this value are the fluctuations and
drifts of the $y$-position of the VDT of $0.5(\pm 0.2)~\mathrm{\mu
m}$~rms (measured by observing the $y$-position of atoms trapped
in the VDT with the ICCD camera). A further contribution to
$\Delta y_\mathrm{insert}$ comes from the finite confinement of
atom 1 in the VDT due to its thermal motion along the
$y$-direction.
\end{thebibliography}
\end{document}